\documentclass[aps,prb,twocolumn,showpacs,floatfix]{revtex4}

\usepackage{graphicx}

\begin{document}

\title{Thin superconductors and SQUIDs in perpendicular
       magnetic field}

\author{Ernst Helmut Brandt}
\affiliation{Max-Planck-Institut f\"ur Metallforschung,
    D-70506 Stuttgart, Germany}

\date{\today}

\begin{abstract}
   It is shown how the static and dynamic electromagnetic
properties can be calculated for thin flat superconducting
films of any shape and size, also multiply connected as used
for SQUIDs, and for any value of the effective magnetic London
penetration depth $\Lambda$. As examples, the distributions of
sheet current and magnetic field are obtained for rectangular
and circular films without and with slits and holes,
in response to an applied perpendicular magnetic field and
to magnetic vortices moving in the film. The self energy and
interaction of vortices with each other and with an applied
magnetic field and/or transport current are given. Due to the
long ranging magnetic stray field, these energies depend on
the size and shape of the film and on the vortex position
even in large films, in contrast to the situation in large
bulk superconductors. The focussing of magnetic flux into
the central hole of square films without and with a radial
slit is compared.
\end{abstract}


\maketitle

\section{Introduction}  

   The calculation of the electromagnetic properties of thin
superconducting films of finite size as used, e.g., in
Superconducting Quantum Interference Devices (SQUIDs) \cite{1}
is a complicated problem since these sensitively depend on
the shape and size of the film. This is so since the currents
in thin films are not screened as in bulk superconductors, but
interact via the magnetic stray field they generate outside
the film. In particular, the self energy of a magnetic vortex
in thin films (called Pearl vortex \cite{2}) and the interaction
between two such vortices depend on their position in the film
even for very large films. This means a vortex is never
``far from the film edges'', in contrast to the behavior of
vortices in the bulk, whose energy, current density, and
magnetic field become independent of the vortex position
and of the specimen shape when the distance from the surface
is much larger than the London penetration depth $\lambda$. This
is so since in the bulk the factor $1/(1 +k^2_\perp\lambda^2)$
in the Fourier transforms (with wave vector components
$k_x$, $k_y$, $k_z$,
$k_\perp^2 = k_x^2 +k_y^2$) causes the fields and currents to
decay exponentially over the length $\lambda$ at large distances
from the vortex core. In thin films of thickness $d < \lambda$ the
effective magnetic penetration depth $\Lambda =\lambda^2/d$ is
larger than $\lambda$ and the Fourier transforms contain a factor
$1/({1\over 2}k_\perp + k_\perp^2 \Lambda)$ that describes also
the long-range non-exponential interaction of vortices via the
magnetic stray field outside the film.

  But even films in the ideal Meissner state containing
no vortices present a difficult problem. Properties of
macroscopic circular disks and rings in a perpendicular
applied magnetic field were calculated recently for ideal
screening ($\Lambda =0$) \cite{3} and for arbitrary
$\lambda$. \cite{4} When this ring has a radial slit, e.g.
in a washer-shaped SQUID, the circular symmetry is lost, but
some properties like the sheet current and the concentration
of magnetic flux into the central hole (flux focussing)
can still be calculated approximately from this circular
symmetric model by forcing the current in the
ring to be zero. \cite{3,4} (This situation may be achieved
by appropriate magnetic history.) Below we shall compare
this approximation with the exact two-dimensional (2D)
solution for a slitted ring and find partial agreement
(Sec.~III).

  While the slitted ring or slitted square film with an
applied magnetic field $H_a$ and/or transport current $I_a$
are simply connected geometries (Fig.~1, right two plots),
a closed ring or a slitted film with the entrance of the
slit short-cut by superconducting contacts (e.g. by weak
links) present multiply connected geometries
(Fig.~1, left two plots). These are more difficult
to calculate since the (quantized) magnetic flux $\Phi$
(or fluxoid $\Phi_{\rm f}$ when $\Lambda>0$) trapped in
this hole, and the current $I$ circulating around the hole,
are additional parameters, besides $H_a$ and $I_a$.
In films with $n$ holes or slots that are fully surrounded
by superconducting material, there are $n$ such fluxoids
$\Phi_{{\rm f}i}$ and currents $I_i$ that depend on the magnetic
history and that may be used to define $n$ self-inductances
$L_i=\Phi_{{\rm f}i} / I_i$ and $n(n-1)/2$ mutual inductances
$M_{ij} = \Phi_{{\rm f}i} / I_j$.

  This paper shows how all these (actually 3D) thin-film
problems can be solved numerically by a 2D
matrix inversion method allowing for non-equidistant
grid points. The presented general
equations and concrete examples generalize previous methods
that either work only for equidistant spatial grids \cite{5}
(which are not very accurate near the film edges or
near a narrow slit or small hole), or for general grid did
not account for finite penetration depth
$\Lambda >0$, \cite{6} or assumed simply connected
geometry, \cite{5,6} or applied only to circular disks
or rings. \cite{3,4} In this paper I consider the
electrodynamics of finite-sized macroscopic films that can be
described by London theory, which applies when the magnetic
field is much smaller than the upper critical field $H_{c2}$
and the superconductor is much larger than the coherence
length $\xi$. The application to SQUIDs will be dealt with
elsewhere. \cite{7} In the present paper I shall thus not
need the notions of fluxoid quantization, phase of the
superconducting order parameter, vector potential, and
voltages caused by the Josephson effect, but some
quantities computed here will be needed in the theory of
SQUIDs, e.g., self-inductance and effective area.

  Our 2D matrix inversion method can be quite accurate even
when the number of grid points is not very large. For example,
in the ideal Meissner state with $\Lambda=0$, the computed
current density exactly yields $H_z(x,y,0)=0$ inside the film
($z=0$ is the film plane). For $\Lambda > 0$, accurate results
are obtained even when $\Lambda$ is smaller than the
spacing of our rectangular grid. For similar calculations
using a finite-element method and a different integral kernel
(or matrix) see Ref.~\onlinecite{8}. Analytical \cite{9,4}
and numerical \cite{10} London calculations
were performed in the limit of large $\Lambda$, i.e., for
small disks with radius $R \ll\Lambda$.
(Here I shall not list numerous recent work on mesoscopically
small superconductors with vortices computed from
Ginzburg-Landau theory.)
An elegant and fast method that computes the currents in films
from the magnetic field pattern measured at the film surface
with high resolution, without having to store or explicitly
invert a large matrix, is described by Wijngaarden
et al.; \cite{11,12}  this method has all advantages of the
direct matrix inversion method and avoids the inversion by
Fourier transform, that would require knowledge of the
magnetic field pattern also outside the film area.
The static Bean model for thin films of any shape is computed
by Prigozhin using a variational method. \cite{13}

\section{Calculation method}  

   This section describes how for thin flat superconducting films
of any shape, also multiply connected as needed for SQUIDs, one
can calculate the static and dynamic response to an applied
magnetic field, applied electric current, and to vortices moving
in the film. In such problems the central physical quantity is
the thickness-integrated current density, called sheet
current ${\bf J}(x,y) = \int\! dz\, {\bf j}(x,y,z) = (J_x,\,J_y)$.
For films with constant thickness $d$ and nearly $z$-independent
current density ${\bf j}(x,y,z)$ one approximately has
${\bf J} = {\bf j}d$, but the following equations are more
general, applying also to films with spatially varying thickness
$d(x,y)$ if the typical length of this variation is $\gg d$.

\subsection{Properties of the stream function}  

   Since one has $\nabla \cdot {\bf J} =0$ in the film except
at small contacts, one can express ${\bf J}$ in terms of a
scalar potential or stream function
$g(x,y)$ as ${\bf J} = -{\bf\hat z} \times \nabla g =
 \nabla \times ({\bf\hat z}g) = ( \partial g / \partial y,\,
 -\partial g / \partial x )$. The function $g(x,y)$ has
several  useful properties and interpretations:

 1.~$g(x,y)$ is the local magnetization or density of tiny
 current loops.

 2.~The contour lines of $g(x,y)$ are the current stream lines.
 Typical $g(x,y)$ look like a mountain (Fig.~2).

 3.~ On the boundary of the film one may put $g(x,y)=0$ since
 the boundary coincides with a stream line.

 4. The integral of $g(x,y)$ over the film area equals the
 magnetic moment of the  film if $g=0$ on its edge.

 5.~The difference $g({\bf r}_1) - g({\bf r}_2)$ is the current
 that crosses any line connecting the two points ${\bf r}_1$
 and ${\bf r}_2$.

 6.~If the film contains an isolated hole or slot such that
 magnetic flux can be trapped in it or a current $I$ can circulate
 around it, then in this hole one has $g(x,y) = {\rm const} = I$
 if $g(x,y)=0$ is chosen outside the film.

 7.~In a multiply connected film with $n$ holes, $n$ independent
 constants $g_1$ $\dots$ $g_n$ can be chosen for the values
 of $g(x,y)$ in each of these holes. The current flowing between
 hole 1 and hole 2 is then $g_1 - g_2$.

 8.~A vortex with flux $\Phi_0$ in the film moves in the
 potential $V=-\Phi_0 g(x,y)$,
 since the Lorentz force on a vortex is $-{\bf J}\times {\bf\hat z}
 \Phi_0 = -\Phi_0 {\bf\hat z} \times ({\bf\hat z} \times \nabla g) =
 \Phi_0 \nabla g(x,y) = -\nabla V$.

 9.~A vortex moving from the edge of the film into a hole connected
 to the outside by a slit, at each
 position $(x,y)$ couples a fluxoid $g(x,y)\Phi_0/I$ into this
 hole,  where $g(x,y)$ is the solution that has $g(x,y)=I$ in this
 hole (with closed slit, see point 6) and $g=0$ outside the film.

 10.~When the film has $n$ holes, a vortex in the film at $(x,y)$
 couples a fluxoid $g_i(x,y)\Phi_0/I$ into the $i$th hole (if this
 is connected to the edge by a slit), where
 $g_i(x,y)$ is the solution exhibiting $g_i=I$ in this hole
 and $g_i=0$ in all other (isolated) holes and on the film edge.

\subsection{Amp\`ere's law for thin films}  

   From Amp\`ere's local 3D law ${\bf j} = \nabla \times {\bf H}$
one obtains for a current-carrying film in the plane $z=0$ a
nonlocal 2D relation between the perpendicular magnetic field
$H_z(x,y,0)$ and the stream function $g(x,y)$:
   \begin{equation}  
   H_z({\bf r}) = H_a({\bf r}) + \int_S\! d^2r'\, Q({\bf r},
   {\bf r}') \, g({\bf r}') \,.
   \end{equation}
Here ${\bf r} = (x,y)$, $H_a({\bf r})$ is the $z$ component of the
applied magnetic field, and $S$ is the area of the film. The
integral kernel
 $Q({\bf r}, {\bf r}')$ has the meaning of the magnetic field
(along $z$) caused at point ${\bf r}$ in the plane of the film by a
magnetic dipole (or tiny current loop) of unit strength positioned
at ${\bf r}'$ and directed along $z$. From the known dipole field
in any plane $z =$ const, one obtains formally:
   \begin{equation}  
   Q({\bf r}, {\bf r}') = Q(\rho) =\lim_{z\to 0}\, {2z^2
   -\rho^2 \over 4\pi (z^2 +\rho^2)^{5/2} }\,.
   \end{equation}
Note that this kernel depends only on the distance
$\rho=|{\bf r}-{\bf r}'|$. For $\rho \ne 0$ one has
$Q( \rho ) = -1/(4\pi \rho^3)$, but the
integral of $Q( \rho )$ over the infinite plane vanishes, i.e.,
the total magnetic flux of a dipole is zero in any plane
$z =$ const, also for $z \to 0$, thus $Q(\rho)$ is {\it highly
singular} at $\rho =0$. For explicit calculations one
has to decide how to deal with this singularity of $Q$.

  For numerics, one has to write the integral as a sum.
Defining a 2D grid whose $N$ points ${\bf r}_i$ fill the film
area $S$, one may approximate any integral over $S$ by a sum,
   \begin{equation}  
   \int_S\! d^2r\, f({\bf r}) \approx \sum_{i=1}^N w_i
   f({\bf r}_i) \,,
   \end{equation}
where the $w_i$ are weights, e.g., $w_i = {\rm const} =S/N$ for
equidistant grids that avoid the film edges, see Appendix A.
The accuracy of this numerical integration can be strongly
increased by choosing an appropriate non-equidistant grid,
e.g., a grid that is denser near the boundary of $S$ or near
possible poles or jumps of the integrand $f(x,y)$.
Equation (1) now becomes
   \begin{equation}  
   H_z({\bf r}_i) = H_a({\bf r}_i) + \sum_j Q_{ij}\, w_j\,
   g({\bf r}_j)
   \end{equation}
with the matrix $Q_{ij} = Q({\bf r}_i, {\bf r}_j)$. Equation (4)
is formally solved by matrix inversion, i.e., by writing
   \begin{equation}  
   g({\bf r}_i) = \sum_j K_{ij}\, \big[ H_z({\bf r}_j) -
   H_a({\bf r}_j) \big] \,,
   \end{equation}
where $K_{ij}$ is the inverse matrix
   \begin{equation}  
   K_{ij} = \big( Q_{ij}\, w_j \big)^{-1}
   \end{equation}
defined by the equation
$\sum_l K_{il} (Q_{lj} w_j ) = \delta_{ij}$ with
$\delta_{ij} = 1$ for $i=j$ and $\delta_{ij} = 0$ for $i \ne j$.
Note that, in contrast to $Q_{ij}$, the inverse matrix $K_{ij}$
depends on the shape of the film and not only on the
difference  ${\bf r}_i-{\bf r}_j$. For the film shapes we have
tested, all the matrix elements $K_{ij}$ are found to be
negative, with a sharp negative peak at $i=j$, and tending to
zero when ${\bf r}_i$ or ${\bf r}_j$ approaches the film edge,
see Fig.~3 for an example.
  In principle, the inverse kernel $K(x,y; x',y')$, integrated
over $y$ and $y'$ (the strip width) was introduced \cite{14} and
depicted \cite{15} earlier in the context of the magnetostatic
energy of a tilted and curved narrow superconducting strip with
pinned vortices. Then, and later,\cite{16} $K(x,y; x',y')$
was calculated by iterating an integral equation.

   A useful expression
for the matrix $Q_{ij}$ was obtained in terms of a Fourier
series in Ref.~\onlinecite{5}, but this method works only for
equidistant grids, while for non-equidistant grids the matrix
inversion is singular. This numerical form of $Q_{ij}$ is
thus not very accurate when one is interested in the sharply
peaked ${\bf J}$ and $H_z$ near the film edges, or when the
film exhibits fine structures, e.g., a small hole or narrow
slit, since the maximum number of grid points on present
Personal Computers is limited to about $N=5000$, yielding a
very large $N \times N$ matrix $Q_{ij}$.

   A matrix $Q_{ij}$ that works well for any grid ${\bf r}_i$,
also with non-constant weights $w_i$, is obtained as follows.
From Eq.~(2) one has for $i\ne j$:
   \begin{equation}  
   Q_{i \ne j} = - {1 \over 4\pi |{\bf r}_i-{\bf r}_j|^3 }
   = -q_{ij} \,.
   \end{equation}
The diagonal terms $Q_{ii}$ are obtained from the condition
that the integral of $Q(\rho)$ taken over the infinite area
has to vanish. Splitting this integral into the integral over
the film area $S$ plus the integral over the infinite area
$\bar S$ outside $S$, and writing the first integral as a
sum, we get:
   \begin{eqnarray}  
   0 &=& \int_\infty \! d^2r'\, Q({\bf r}_i-{\bf r}')
   \nonumber \\ &\approx&
   \sum_j Q_{ij} w_j + \!\int_{\bar S} d^2r'
   Q({\bf r}_i - {\bf r}') \,.
   \end{eqnarray}
Defining a 2D function $C({\bf r})$ as an integral over
the film area or as a contour integral over the film edge,
   \begin{equation}  
   C({\bf r})  = \int_{\bar S}  { d^2r' \over 4\pi
   |{\bf r}-{\bf r}'|^3 } = \int_0^{2\pi}\!\!\!
   { d\phi \over 4\pi R(\phi) }
   \end{equation}
with $R = |{\bf r}-{\bf r}'|$ ($\phi$ is the angle between
the vector ${\bf R} = {\bf r}-{\bf r}'$ pointing to the
point ${\bf r}'$ on the film edge  and any fixed
direction, say, the $+x$ axis), and writing
$C_i=C({\bf r}_i)$, $q_{ij} = 1 / ( 4\pi
   |{\bf r}_i-{\bf r}_j|^3 ) $, we get from (8) the
diagonal term $Q_{ii} w_i =\sum_{j\ne i} q_{ij}w_j + C_i$.
The full matrix reads thus:
   \begin{equation}  
   Q_{ij} = (\delta_{ij} -1)\, q_{ij} +\delta_{ij} \Big(
   \sum_{l\ne i} q_{il}\, w_l + C_i\Big)/w_j \,.
   \end{equation}
Note that the terms in (10) should not be rearranged since
$q_{ii} = \infty$. The matrix (10) is well behaved during
inversion, so one may write
   \begin{equation}  
   K_{ij} = \Big[(\delta_{ij} -1)\, q_{ij} w_j +\delta_{ij}
   \Big( \sum_{l\ne i} q_{il}\, w_l +C_i\Big) \Big]^{-1} .
   \end{equation}

  For a rectangular film filling the area $|x| \le a$,
$|y| \le b$ one has explicitly from Eq.~(9):\cite{5,16}
   \begin{equation}  
   C(x,y)  = {1 \over 4\pi} \sum_{p,q}\Big[ (a-px)^{-2}
   + (b-qy)^{-2} \Big] ^{1/2}
   \end{equation}
with $p,q = \pm 1$. Interestingly, expression (12) may be
used also for films that have a hole or slit, or several
holes, or that do not fill the rectangle completely, e.g.,
a circular disk with radius $\le a=b$. In such cases one
has to omit in Eq.~(4) and (5) the grid points that fall
outside the film (but keep the points in isolated holes,
 see Sct.~II E). Therefore, Eq.~(5) with
the explicit kernel $K_{ij}$ from Eq.~(11) and $C_i$ from
Eq.~(12), allows one to compute the stream function
$g(x,y)$ and thus the sheet current  for thin films of
arbitrary shape when one knows the magnetic field
 $H_z(x,y) - H_a(x,y)$ generated inside the film by this
current. Information on $H_z$ outside the film is
{\it not} required for this stable inversion method.

\subsection{Static solution of London equation}  

   In superconductor films with thickness $d < \lambda$,
the London penetration depth, the 3D static London equation
$\lambda^2 \nabla \times {\bf j} + {\bf H} =0$ may be
integrated over the film thickness $d$ to yield the 2D equation
   \begin{equation}  
   H_z(x,y) = -\Lambda [ \nabla \times {\bf J}(x,y) \,]
   \, {\bf\hat z} = \Lambda \nabla^2 g(x,y) \,,
   \end{equation}
where $\Lambda = \lambda^2 / d$ is the effective London
depth of the film. Eliminating $H_z(x,y)$ from Eqs.~(1) and
(13) one obtains an implicit equation for $g(x,y)$:
   \begin{eqnarray}  
   H_a({\bf r}) = -\!\int_S\! d^2r'\, Q({\bf r},{\bf r}') \,
    g({\bf r}') +\Lambda \nabla^2 g({\bf r})  \nonumber \\
        = -\!\int_S\! d^2r'\big[ Q({\bf r},{\bf r}')
       -\delta({\bf r}-{\bf r}')\Lambda \nabla^2 \big]
        g({\bf r}')
   \end{eqnarray}
or with the discretized Eq.~(4),
   \begin{equation}  
   H_a({\bf r}_i) = -\sum_j \big(\, Q_{ij}\, w_j
   - \Lambda \nabla^2_{ij} \,\big)\,  g({\bf r}_j) \,.
   \end{equation}
In it the matrix $\nabla_{ij}^2$ computes the Laplacean
$\nabla^2 =\partial^2/\partial x^2 +\partial^2/\partial y^2$
at ${\bf r} ={\bf r}_i$ of a function defined on a grid, e.g.,
from the values $g({\bf r}_j)$ at ${\bf r}_j ={\bf r}_i$ and
its four nearest neighbors. Equation (15) is solved
for $g(x,y)$ by matrix inversion:
   \begin{equation}  
   g({\bf r}_i) = - \sum_j K_{ij}^\Lambda\, H_a({\bf r}_j)
   \end{equation}
with the inverse matrix
   \begin{equation}  
   K_{ij}^\Lambda = \big( Q_{ij}\, w_j
    - \Lambda \nabla_{ij}^2  \big)^{-1}
   \end{equation}
now depending on $\Lambda$. This matrix inversion is the
more stable the larger is $\Lambda$, since finite $\Lambda$
increases the diagonal terms and makes the resulting
$K_{ij}^\Lambda$ a smoother function as compared to the
case $\Lambda=0$ considered in Eq.~(6).
Examples for $H_z(x,y)$ are shown in Fig.~4 for a square
with slit and hole, while Fig.~5 shows some profiles
$H_z(x,0)$ along the $y$-axis for the same square.
Note that even small $\Lambda = 0.01 a$ allows $H_z$ to
partly penetrate the entire film.

\subsection{Dynamic solution of London equation}  

   The time dependent behavior of superconducting films
containing vortices may be described within continuum theory
by the following  realistic relation between the local electric
field ${\bf E}(x,y,t)$ and the sheet current ${\bf J}$ and
magnetic induction ${\bf B} = \mu_0 {\bf H}$: \cite{5}
   \begin{equation}  
   {\bf E} = \rho_s(J,B)\, {\bf J}({\bf r},t) + \mu_0\Lambda
   \,{\bf\dot J}({\bf r},t) \,.
   \end{equation}
Here $\rho_s = \rho/d$ is the sheet resistivity caused by moving
vortices, and the second term with $\Lambda = \lambda^2 /d$
and  ${\bf\dot J}= \partial{\bf J}/\partial t$ is
the London term describing acceleration of Cooper pairs. The
isotropic model (18) assumes that the resistivity $\rho$
depends only on the magnitudes $J$ and $B$. For example,
without vortex pinning and Hall effect, one has free flux flow
with $\rho = \rho_{\rm FF} \approx (B/B_{c2}) \rho_n$, where
$\rho_n$ is the resistivity in the normal conducting state and
$B_{c2}$ is the upper critical field. For thermally activated
depinning a realistic model is $\rho =\rho_0 | J/J_c(B) |^\sigma$
with creep exponent $\sigma \gg 1$ and an in general $B$ dependent
critical sheet current $J_c(B)$. For a generalization to
anisotropic superconductors see Ref.~\onlinecite{5}.

  From the induction law ${\bf\dot B} = -\nabla\times {\bf E}$,
which in the film plane reduces to
$\dot B_z =\partial E_x /\partial y -\partial E_y /\partial x$,
and from   ${\bf J} = -{\bf\hat z} \times \nabla g$, one obtains
 $\mu_0 \dot H_z = \dot B_z = \nabla [\rho_s \nabla g ]
 +\mu_0 \Lambda \nabla^2 \dot g$. Inserting this into the time
derivative of Eq.~(1) one finds an equation for $g({\bf r},t)$:
   \begin{eqnarray}  
   \int_S\! d^2r'\big[\, Q({\bf r},{\bf r}')
   -\delta({\bf r}-{\bf r}')\Lambda \nabla^2 \,\big]
   \dot g({\bf r}',t)                     \nonumber \\[-1mm]
  =f({\bf r},t) -\dot H_a({\bf r},t) \,,  \nonumber \\[ 2mm]
   f({\bf r},t) =  \mu_0^{-1} \nabla
     \big[\, \rho_s({\bf r},t) \nabla g({\bf r},t) \,\big] \,.
   \end{eqnarray}
In discretized form this becomes [cf.\ Eq.~(15)]:
   \begin{equation}  
   \sum_j\! \big( Q_{ij}\, w_j -\Lambda \nabla^2_{ij} \big)\,
   \dot g({\bf r}_j) =f({\bf r}_i,t) -\dot H_a({\bf r}_i,t) \,.
   \end{equation}
Inverting this one obtains the equation of motion for $g(x,y,t)$
in explicit form:
   \begin{equation}  
   \dot g({\bf r}_i,t) = \sum_j K_{ij}^\Lambda \big[\,
    f({\bf r}_j,t) - \dot H_a({\bf r}_j,t) \,\big]
   \end{equation}
with $K_{ij}^\Lambda$ from Eq.~(17). In the Meissner state or
for rigidly pinned vortices one has $\rho_s = 0$ and these
dynamic equations reduce to the static equations of Sct.~II C.

\subsection{Multiply connected films}  

  Multiply connected films have one or more holes or slots that
are completely surrounded by superconducting material and thus
can trap magnetic flux. As a fundamental example I consider here
a film containing one such hole with flux trapped such that a
current $I$ circulates around the hole when no magnetic field is
applied, $H_a=0$. In this case one has $g(x,y)=0$ outside the
film and $g(x,y)=I$ in the hole, and inside the film $g(x,y)$
smoothly goes from 0 to $I$.  Generalization of this example
to the presence of more holes and to $H_a \ne 0$ is possible by
linear superposition.

   This problem may be solved in three steps. First, consider
the situation where $g=I$ in the hole and $g=0$ everywhere outside
the hole. This means a sharply localized sheet current of size $I$
flows along the edge of the hole where this $g(x,y)$ has a jump.
Such an edge current formally can be caused by an effective
applied field
   \begin{equation}  
   H_a^{\rm eff}({\bf r}_i) = -I\!\! \sum_{j {\rm~in~ hole}} \!\!
     \big(\, Q_{ij}\, w_j - \Lambda \nabla^2_{ij} \,\big) \,,
   \end{equation}
cf.~Eq.~(15). Next the real sheet current in the film is found as the
 ${\bf J} = -{\bf\hat z} \times \nabla g$ that generates this
 $H_a^{\rm eff}({\bf r})$ inside the film, cf.~Eq.~(16):
   \begin{eqnarray}  
   g({\bf r}_i) =& -\!\!{\displaystyle \sum_{j {\rm~in~film}}}\!\!
                    K_{ij}^\Lambda\, H_a^{\rm eff}({\bf r}_j)
     &~{\rm for~} {\bf r}_i {\rm ~in~the~film},~~~\nonumber\\[-1mm]
   g({\bf r}_i) =& I
     &~{\rm for~} {\bf r}_i {\rm ~in~the~hole},~~~\nonumber\\[ 1mm]
   g({\bf r})~  =& 0
     &~{\rm outside~the~film}.~~~
   \end{eqnarray}
Finally, the real magnetic field in the entire plane $z=0$ is
obtained as the field caused by this current, cf.~Eq.~(4):
   \begin{equation}  
   H_z({\bf r}_i) =  \sum_j Q_{ij}\, w_j\, g({\bf r}_j) \,,
   \end{equation}
where now the sum is over all ${\bf r}_j$ in the film and  in
the hole. This method works for any value of $\Lambda$ and
yields continuous functions $g(x,y)$ and $H_z(x,y)$ when
$\Lambda > 0$. When $\Lambda = 0$ (ideal screening) the resulting
$H_z(x,y)$ has sharp jumps at the film edges since in that case
$H_z$ exactly vanishes inside the film and has sharp
infinities just outside the film edges (also in the hole), where
$H_z \propto \delta^{-1/2}$ ($\delta$ is the distance from
the edge). The current density in the film for $\Lambda=0$
diverges similarly, but on the inner side of the edges, where
$g \propto \delta^{1/2}$ and $J \propto \delta^{-1/2}$.
For $\Lambda >0$ the sheet current at the edges is finite and
the infinities of $H_z$ are logarithmic on both sides of all
edges.
  \begin{figure}   
\includegraphics[scale=.48]{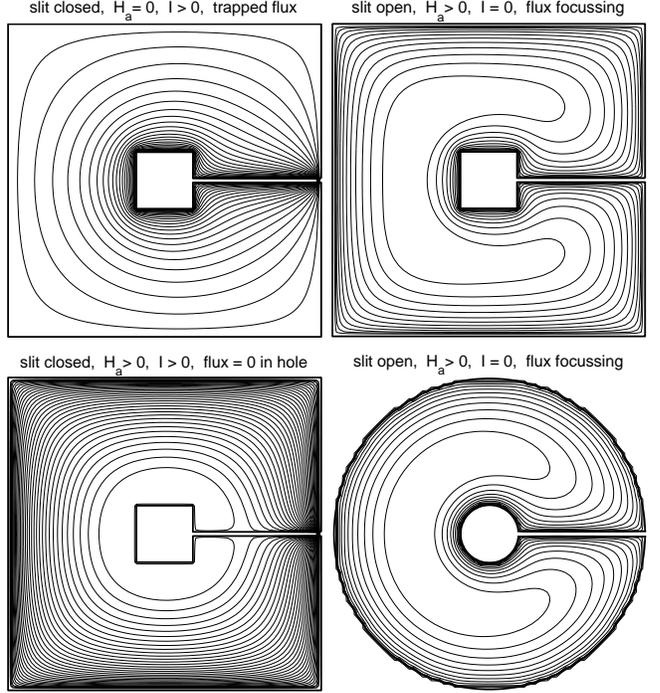}
\caption{\label{fig1}  Current stream lines in a thin film square
with square hole and radial slit, and in a circular disk with
circular hole and slit, in the ideal Meissner state $\Lambda=0$.
Top left: Slit bridged at the edge, circulating current $I >0$
flows due to flux trapped in the hole and slit, no applied field
$H_a=0$. Top right and bottom right:
                 Slit open, applied field $H_a > 0$, magnetic flux
enters the slit and is focussed into the hole where $H(x,y) > H_a$.
Bottom left: Closed slit, applied field $H_a >0$, some current $I>0$
flows such that the flux in hole and slit is exactly zero (ideal
screening); this state is a superposition of the two upper states.
Note that the current near the hole circulates in opposite
direction, except in the trapped-flux case.
}
  \end{figure}    
  \begin{figure}   
\includegraphics[scale=.74]{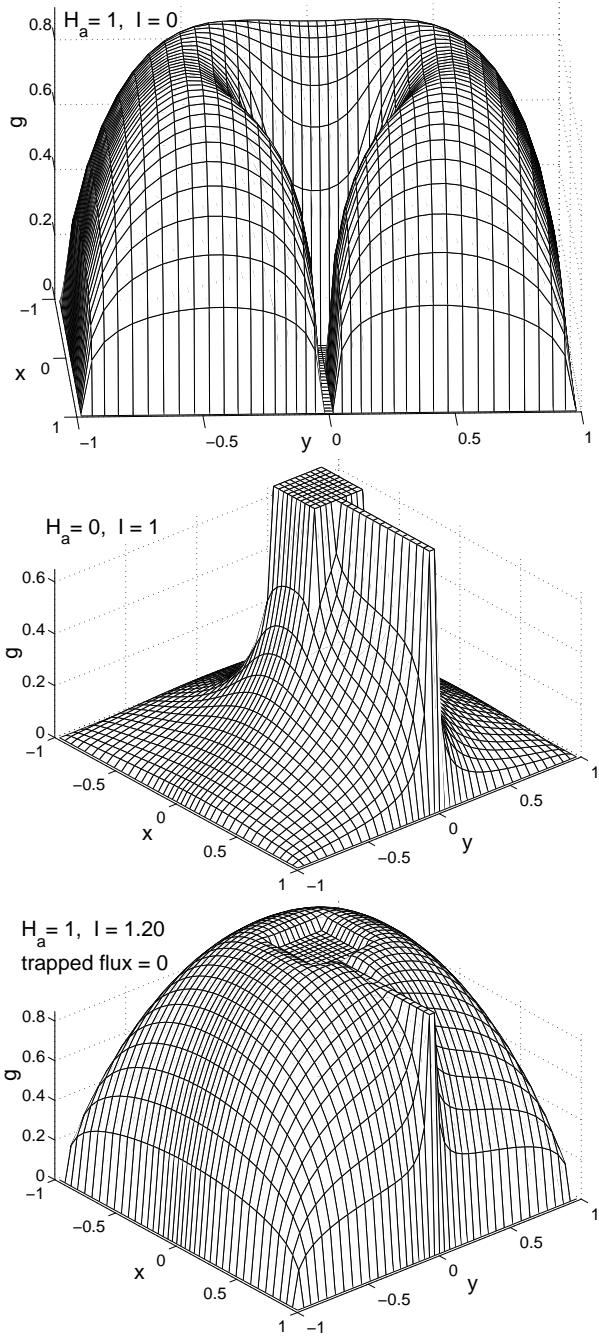} \vspace{-1mm}
\caption{\label{fig2}  The three examples of Fig.~1 for the
stream function $g(x,y)$ in a square film (size $2 \times 2$ in
units of half width $a$) with square hole (half width $a_1=0.2 a$)
and open slit (width 0.04 a), for
$40 \times 40$ grid points, as  Fig.~3. For penetration
depth $\Lambda = 0$ (ideal screening).
At the film edges $g(x,y)$ goes to zero with vertical slope,
and outside the film $g=0$ (for $\Lambda >0$, the slope of $g$
at the edges is finite, equal to the sheet current).
Top: Constant applied field $H_a=1$, open slit meaning a current
$I=0$, thus $g=0$ in slit and hole (like Fig.~1 top right).
Middle: $H_a=0$, current $I=1$ flowing around the hole and closed
slit due to trapped flux, yielding $g=1$ in hole and slit
(like Fig.~1 top left).
Bottom: Top and middle cases superimposed such that the flux
trapped in the hole and slit is zero (like Fig.~1 bottom left):
weights 1 and 1.2, thus in slit and hole $g=I=1.2 aH_a$.
  }
  \end{figure}    
  \begin{figure}   
\includegraphics[scale=.48]{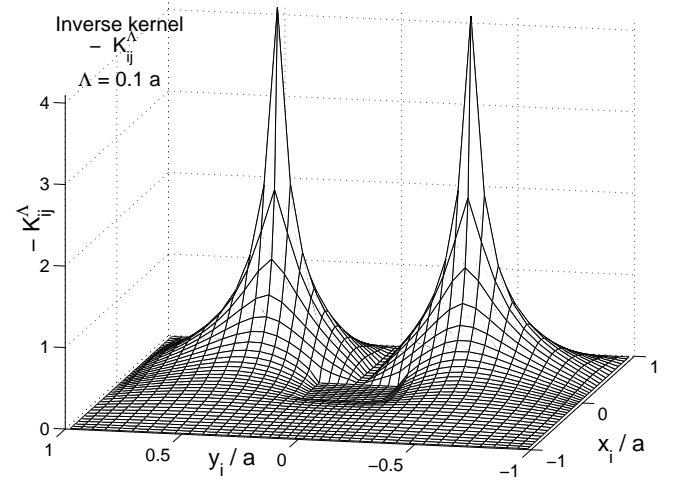}
\caption{\label{fig3} Example for the inverse matrix
$K_{ij}^\Lambda = K^\Lambda({\bf r}_i, {\bf r}_j)$ in a square film
(half width $a$) with square hole (half width $a_1 = 0.2 a$)
and open slit (width $0.04 a$), similar to the squares in Figs.~1 and
Fig.~2.  For $40\times 40$ grid points ${\bf r}_i$ and constant
 ${\bf r}_j = (0.48, 0.42)a$, penetration depth $\Lambda = 0.1 a$.
At the film edges and in hole and slit one has $K_{ij}^\Lambda=0$.
The plotted $K_{ij}^\Lambda$ is approximately symmetric in $i,j$ since
here the weights $w_i \approx $ const. It also shows the interaction
between a vortex at $(x,y)$ and a vortex pair (due to the imposed
symmetry) sitting at $(x_j,y_j)=(0.48, \pm 0.42)a$, cf.\ Eq.~(29).
Its contours are the current stream lines of this vortex pair.
For $\Lambda =0$,  $K_{ij}^\Lambda =K_{ij}$ looks similar, but the
peak is higher and the slopes at the edges are vertical.
  }
  \end{figure}    
  \begin{figure}   
\includegraphics[scale=.54]{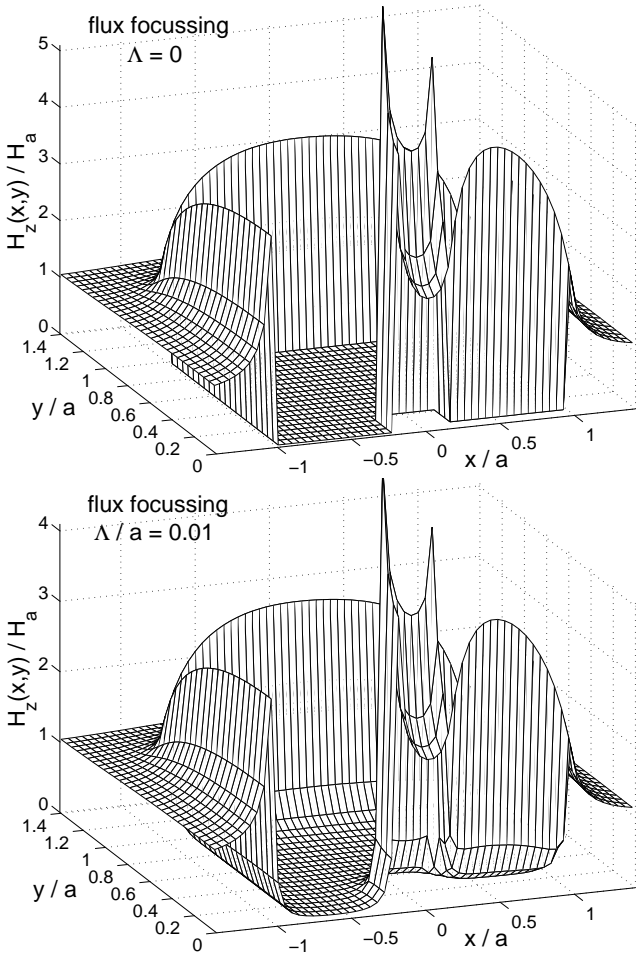}
\caption{\label{fig4}  The magnetic field $H_z(x,y)$ in the plane of
a thin square with hole and slit, $H_a>0$, $I=0$ (case of flux
focussing, see Fig.~1 top right and Fig.~2 top) for ideal screening
$\Lambda=0$ (top) and for $\Lambda/a=0.01$ (bottom) ($60 \times 60$
grid points, only half the square is shown). Note
that even such small $\Lambda$ strongly changes $H_z(x,y)$ inside
the superconductor, which penetrates much farther than $\Lambda$.
The corresponding profiles $H_z(x,0)$ are show in Fig.~5.
  }
  \end{figure}    
  \begin{figure}   
\includegraphics[scale=.48]{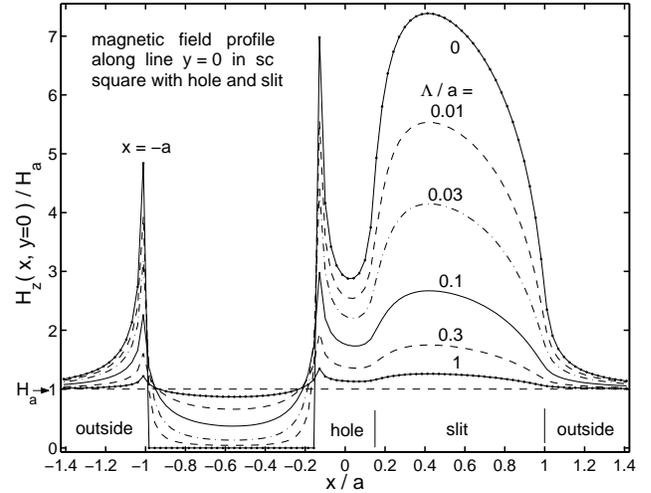}
\caption{\label{fig5}  The profiles of the magnetic field $H_z(x,0)$
in the thin square of Fig.~1 (top right) taken along the $x$ axis
that passes through the hole and slit, in units of the applied field
$H_a$ and shown for several values of the 2D magnetic penetration
depth $\Lambda = \lambda^2 /d = 0$, 0.01, 0.03, 0.1, 0.3, and 1 in
units of the half width $a$ of the square. Same case as in Fig.~4
but with more grid points ($100 \times 100$). Note that in the center
of the square hole the magnetic field is enhanced by a factor of 3
when $\Lambda=0$, $H_z(0,0) \approx 3 H_a$ (flux focussing); in the
narrow slit $H_z(0.4a, 0) = 7.4 H_a$ is even higher. Finite $\Lambda$
reduces this enhancement and the spatial variation of $H(x,0)$.
  }
  \end{figure}    
  \begin{figure}   
\includegraphics[scale=.52]{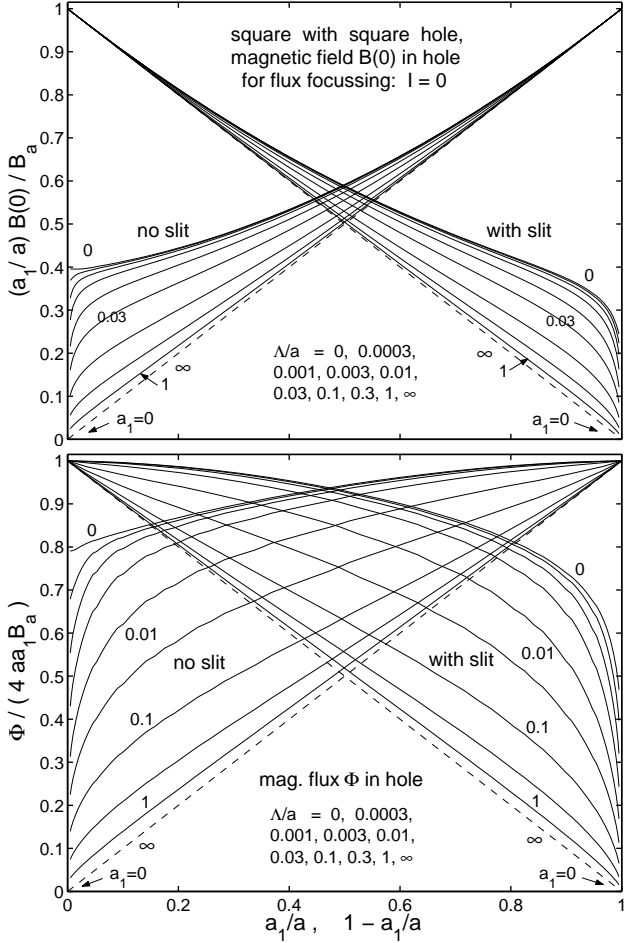}
\caption{\label{fig6} Thin superconducting square (half width $a$)
with central square hole (half width $a_1$) in applied field
$B_a =\mu_0 H_a >0$ and with no circulating current
($I=0$, case of flux focussing). Plotted are the magnetic field
$B(0) = \mu_0 H_z(0,0)$ in the center of the hole (top)
and the magnetic flux $\Phi$ inside the hole (bottom), for squares
without slit (left, versus $a_1/a$) and with a narrow radial slit
(right, versus $1-a_1/a$), like in Fig.~1, for several values of
the 2D penetration depths $\Lambda/a = 0 \dots \infty$.
For $\Lambda=0$ and $a_1/a \to 0$, both the minimum field and the
average field in the hole diverge,
$2B(0)/B_a \approx \Phi/4a_1^2 B_a \approx 0.80 a/a_1$, see also
the corresponding Figs.~15 and 16 of Ref.~\onlinecite{4} for rings.
For small holes, $a_1 \ll a$, the presence of a slit reduces this
field enhancement.
  }
  \end{figure}    
  \begin{figure}   
\includegraphics[scale=.47]{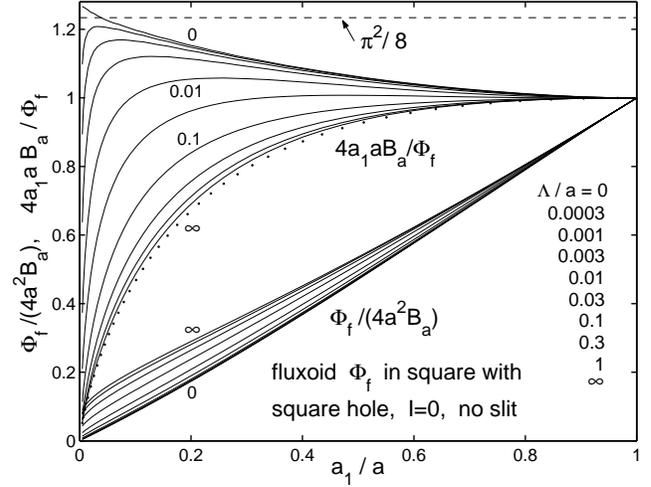}
\caption{\label{fig7} The fluxoid $\Phi_{\rm f}$ trapped in the square
hole of a thin square without radial slit in the flux-focussing
case of Fig.~6. This figure is very similar to Fig.~14 of
Ref.~\onlinecite{4} for rings. For small and large $\Lambda$
nearly the same limiting curves are reached as for rings.
The dots show the large-$\Lambda$ limit for rings with radii
 $a_1$ and $a$: $4a_1a B_a/\Phi_{\rm f} \approx
 2 a_1a \ln(a/a_1) /(a^2 -a_1^2)$.\cite{4} For $\Lambda = 0$ this
 $\Phi_{\rm f}$ coincides with the $\Phi$ in Fig.~6 (bottom left).
  }
  \end{figure}    
  \begin{figure}   
\includegraphics[scale=.52]{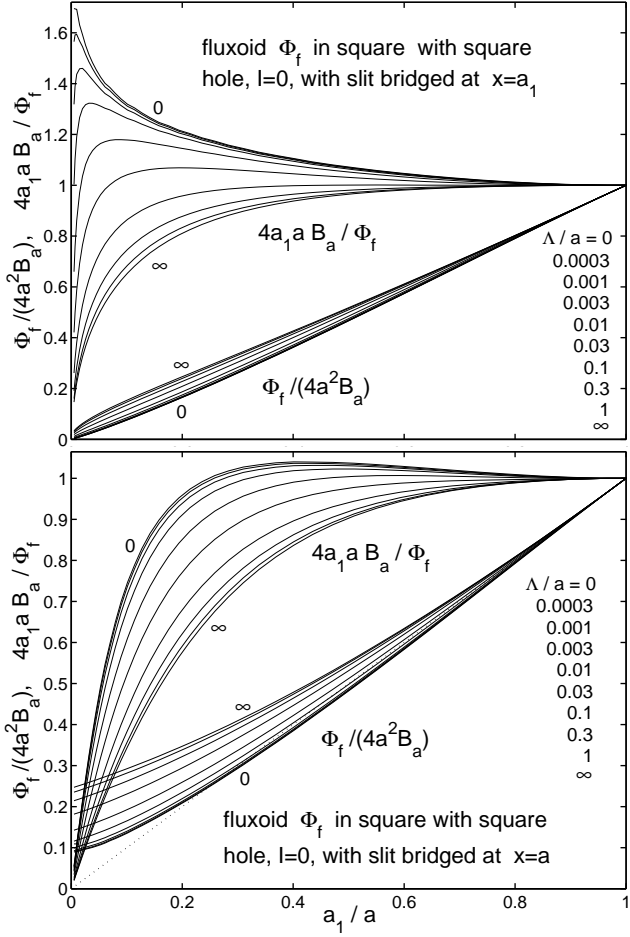}
\caption{\label{fig8} The fluxoid $\Phi_{\rm f}$ trapped in the square
hole of the square of Fig.~7, but now with a narrow radial slit.
Since the integration path of the fluxoid must run inside the
superconductor, the path has to cross the slit by a narrow bridge,
through which no current flows since $I=0$ in this flux-focussing
case. The plot shows the fluxoid when this bridge is chosen at
$x=a_1$ (where the slit enters the hole, top) and at $x=a$ (where
the slit exits the square, bottom). For large holes these two
$\Phi_{\rm f}$ are similar, but for small holes the larger integration
path (bottom) yields a larger fluxoid since the integration includes
the not negligible magnetic flux inside the narrow slit.
For $\Lambda=0$ the fluxoid of the smaller path (top)
coincides with the flux $\Phi$ of Fig.~6 (bottom right).
  }
  \end{figure}    
  \begin{figure}   
\includegraphics[scale=.55]{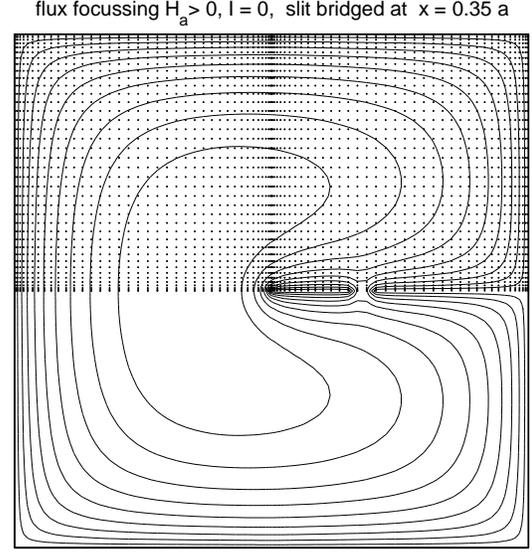}
\caption{\label{fig9} The current stream lines in a thin
superconductor square ($|x|\le a$, $|y|\le a$) with radial slit
running at $y=0$ from $x=0$ to $x=a$, with a superconducting bridge
at $x=a_2=0.35 a$. Shown is the flux-focusing case $H_a>0$, $I=0$,
i.e., no current crosses the bridge. $\Lambda/a = 0.01$. The dots
mark the numerical grid of $80 \times 80$ points used for this plot.
  }
  \end{figure}    
  \begin{figure}   
\includegraphics[scale=.46]{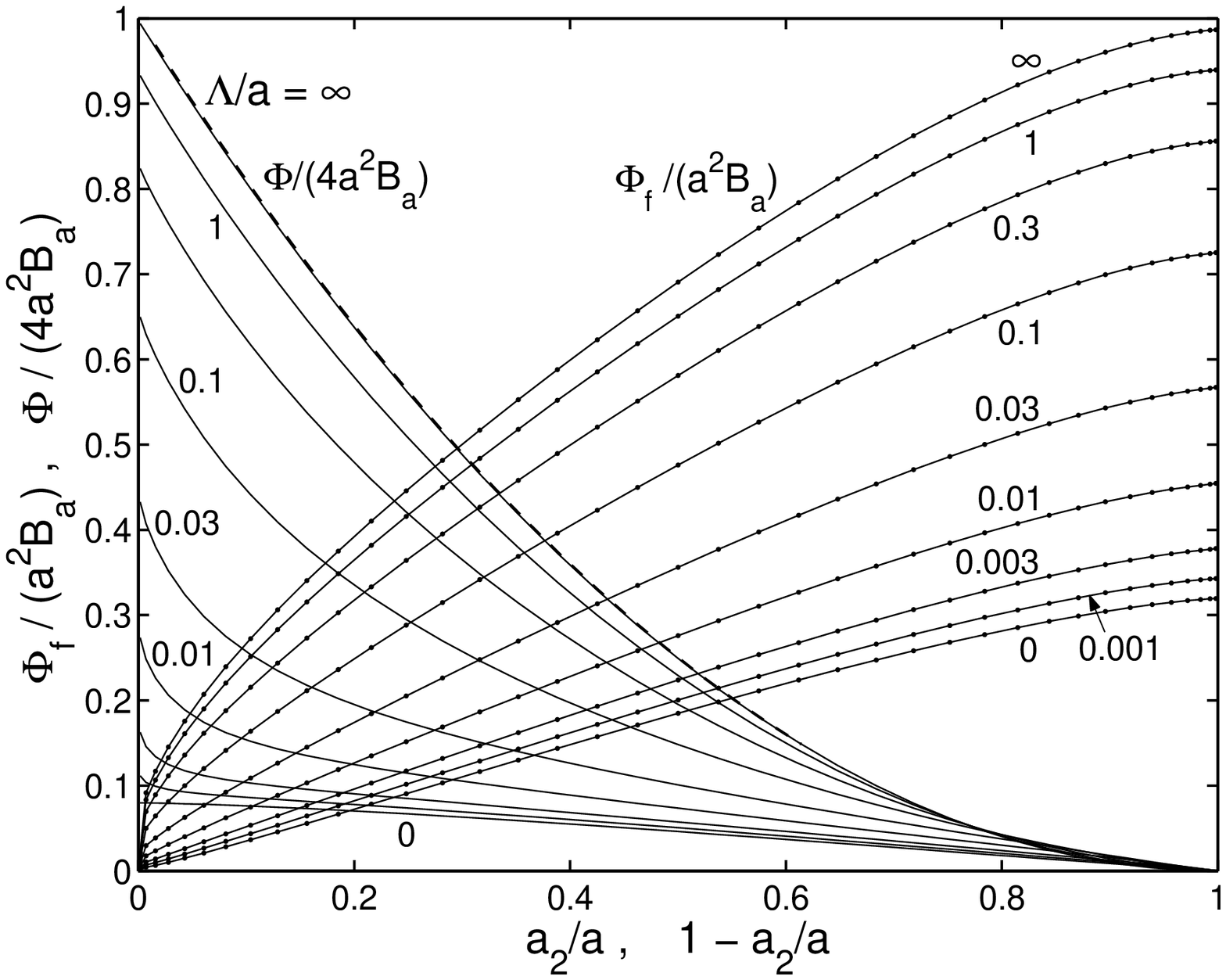}
\caption{\label{fig10} The fluxoid $\Phi_{\rm f}$ and magnetic flux
$\Phi$ inside a concentric square of half width $a_2$ passing through
the bridge (at $x=a_2$) of the slitted square of Fig.~9, plotted
versus $a_2/a$ for several values $\Lambda/a$. For better presentation
$\Phi_{\rm f}$ is shown 4 times larger than $\Phi$, and $\Phi$ is
plotted versus $1-a_2/a$. For $\Lambda=0$ one has $\Phi_{\rm f}=\Phi$.
For $\Lambda/a \gg 1$ one has $\Phi = 4a_2^2 B_a$ (dashed line, full
penetration), and $\Phi_{\rm f} \approx (a_2/a)^{0.6} a^2 B_a$.
  }
  \end{figure}    

\subsection{Individual vortices in the film}  

   The sheet current, magnetic field, and energy of vortices in
the film is obtained by linear superposition from the solution
for one vortex and the interaction energy of a vortex pair.
In a film of finite size all these results explicitly
depend on the vortex positions and not only on their distances
due to the strong effect of the film edges.
The existence of one vortex at position ${\bf r}_v$ modifies the
static London Equation of Sct.~II C to give
$\lambda^2 \nabla \times {\bf j} + {\bf H} = (\Phi_0/\mu_0)
 \, {\bf\hat z}\, \delta_2({\bf r}-{\bf r}_v)$, where
$\Phi_0 = h/2e = 2.07\cdot 10^{-15}$ Tm$^2$ is the quantum of
flux and  $\delta_2({\bf r})$ is the 2D delta function.
Equations (13) and (14) then become
   \begin{eqnarray}  
   H_z({\bf r}) - \Lambda \nabla^2 g({\bf r}) =
    \mu_0^{-1}\Phi_0 \, \delta_2({\bf r}-{\bf r}_v) \,, \\
      \int_S\! d^2r'\big[ Q({\bf r},{\bf r}')
      -\delta({\bf r}-{\bf r}')\Lambda \nabla^2 \big]
      g({\bf r}') \nonumber \\   = \mu_0^{-1}\Phi_0 \,
      \delta_2({\bf r}-{\bf r}_v) -H_a({\bf r}) \,.
   \end{eqnarray}
Writing the integral as a sum yields
   \begin{eqnarray}  
   \sum_j \big(\, Q_{ij}\, w_j
   - \Lambda \nabla^2_{ij} \,\big)\,  g({\bf r}_j) ~~~~~
      \nonumber \\ = \mu_0^{-1}\Phi_0 \,
   \delta_2({\bf r}_i-{\bf r}_v) - H_a({\bf r}_i) \,.
    \end{eqnarray}
To invert this and find $g({\bf r}_i)$ and the vortex interaction,
we have to assume that the vortex sits on a grid point,
 ${\bf r}_v ={\bf r}_j$. Averaging over the grid cell centered at
 ${\bf r}_j$ and having an area $w_j$, replaces
 $\delta_2({\bf r}_i-{\bf r}_j)$ by $\delta_{ij} /w_j$. Inverting
this and performing the sum containing $\delta_{ij}$ then yields
the stream function caused by a vortex positioned at ${\bf r}_j$
and by the applied field $H_a({\bf r}_i)$:
   \begin{eqnarray}  
   g({\bf r}_i) =  \mu_0^{-1}\Phi_0 K_{ij}^\Lambda /w_j \,
     - \sum_l K_{il}^\Lambda  H_a({\bf r}_l)
   \end{eqnarray}
with the inverse matrix $K_{ij}$ from Eq.~(17), see also Eq.~(16).

   A second vortex sitting at ${\bf r}_i$ sees the potential
 $- \Phi_0 g({\bf r}_i)$; the interaction energy between two
vortices positioned at ${\bf r}_i$ and ${\bf r}_j$ is thus:
   \begin{eqnarray}  
   V({\bf r}_i, {\bf r}_j) = V_{ij} = V_{ji} = - \mu_0^{-1}
   \Phi_0^2\,  K_{ij}^\Lambda /w_j \,.
   \end{eqnarray}
This potential is repulsive (positive) and sharply peaked at
${\bf r}_i ={\bf r}_j$, since all the $K_{ij}^\Lambda$ are negative.
One can show that the matrix $K_{ij}^\Lambda /w_j$ is indeed
symmetric in $i,j$,\, $K_{ij}^\Lambda /w_j =K_{ji}^\Lambda /w_i$.
For $\Lambda=0$ this is directly seen from the definition,
Eq.~(6), writing $\delta_{ij} = \sum_l K_{il} (Q_{lj} w_j )
 = \sum_l (K_{il}/w_l) (w_l Q_{lj} w_j ) $, which shows that
 $K_{il}/w_l$ is the inverse of the symmetric matrix
 $w_l Q_{lj} w_j$  and is thus symmetric itself. For
 $\Lambda > 0$ one also has to prove that the operator
 $\nabla_{ij}/w_j$ is symmetric, see Appendix.

 When the film contains several vortices  positioned at some
of the grid points ${\bf r}_j$, then the sum over these
 ${\bf r}_j$ has to be performed in the first term of
Eq.~(28), and the total energy of this vortex system becomes:
   \begin{eqnarray}  
   F &=& \sum_i F_s({\bf r}_i) + \sum_{j>i} V_{ij} + \sum_i
   V_a({\bf r}_i)  \nonumber \\  &\approx& {1\over 2}
   \sum_{j,i} V_{ij} + \sum_i V_a({\bf r}_i) \,,
   \end{eqnarray}
where the sums are over the vortex positions $i,j$,
 $F_s({\bf r}_i) \approx {1\over 2} V_{ii}$ is the self energy
of the vortex, $V_{ij}$ is the vortex interaction (29), and
$V_a({\bf r}_i) = \sum_l K_{il}^\Lambda  H_a({\bf r}_l)$
is the potential caused by the applied field $H_a({\bf r})$.
For constant $H_a$ the external potential
$V_a({\bf r}_i) = H_a \sum_l K_{il}^\Lambda$
has the shape of a negative trough which is zero along the
film edges. The vortices are thus pulled into the film by
this potential.

\subsection{Self energy of a vortex in the film}  

   In contrast to vortices in large bulk superconductors,
the self energy of a vortex in a thin film depends on the film
size and shape and on the vortex position even when it is far
from the film edges. Calculating this vortex energy from the
2D current density in the film and the 3D magnetic stray field
outside the film would be a formidable task. Fortunately,
a much simpler calculation is possible using our above results
and the known Lorentz force $\Phi_0 \nabla g({\bf r})$ on the
vortex:  In a thought experiment, we move a first vortex
from the film edge to its final position ${\bf r}_i$. At the
edge the energy of this vortex, and also the interaction
with other vortices needed later, are zero. Then its energy
increases according to the integrated Lorentz force that
originates from the interaction of this vortex with its own
sheet current (more precisely: with the film edges, or
with its images if an image method can be used, but this
argument will not be required here). When the vortex has
reached position ${\bf r}_i$ its energy is just its
self energy $F_s({\bf r}_i)$. Now move a second vortex
from the edge and merge it with the first one. The
self energy of this new vortex, $4 F_s({\bf r}_i)$, is
composed of the two self energies and the energy needed
to move the second vortex against the sheet current of the
first one, equal to the interaction energy $V_{ii}$,
Eq.~(29). From this we obtain the self energy
$F_s({\bf r}_i) = V_{ii} /(4-2) ={1\over 2} V_{ii}$
used in Eq.~(30).

 This result is exact within our
numerical method, but in real films the self energy depends
on the logarithm of the vortex core radius $\approx \xi$,
the coherence length. In our numerics $\xi$ is effectively
replaced by some cut-off length of the order of the grid
spacing. If required, an improved consideration of the
vortex core is possible if its radius exceeds the local grid
spacing. The core shape may then be taken from the
GL solution for infinite films. \cite{17} If $\xi$ is
smaller than the grid spacing, the correct self energy is
slightly larger than ${1\over 2} V_{ii}$, by a
position-independent constant.

\subsection{The fluxoid in films}  

   The fluxoid $\Phi_{\rm f}$ inside a given closed path $S$
inside the film is defined as the magnetic flux through this
loop plus the 2D penetration depth $\Lambda$ times the path
integral of the sheet current,  $\Phi_{\rm f} /\mu_0 = \int_S
dx\,dy\,H_z(x,y) +\Lambda\oint d{\bf S}\, {\bf J}(x,y)$.
When $g(x,y)$ and $H_z(x,y)$ are given on a rectangular
grid, the integration path is conveniently chosen along
a closed rectangle which runs in the middle between the
grid points (App.~A). The components $J_x$ and $J_y$ are then
obtained from the difference of the values of $g(x,y)$ at
neighboring grid points, while the flux is the sum of
$w_k H_z(x_k,y_k)$ over all points $k$ inside this loop
(App.~A). The fluxoid obtained in this way from our solutions
is indeed independent of the chosen path to within 4 to 5
significant digits, confirming thus that the solutions
$g(x,y)$ and $H_z(x,y)$ are accurate also for finite $\Lambda$
and even when $\Lambda \ll a$, see Figs.~7 and 8 below.
Surprizingly, the obtained $\Lambda$-dependent $g$ and $H_z$
are quite accurate even when $\Lambda$ is much smaller
than the spacing of the numerical grid (typically
$\ge a/50$), down to $\Lambda/a = 10^{-4}$. Besides this,
the solutions for $\Lambda=0$ are very accurate, with
$H_z=0$ inside the film, see Fig.~5.

\section{Example: Flux Focussing}  

   To illustrate how this method works and to check previous
approximations of a slitted ring by a full ring, \cite{3,4}
I discuss here the case of flux focussing in some detail.
Consider a thin square film of half width $a$ with a
central square hole of half width $a_1$ ($a_1  \le |x| \le a$,
$a_1  \le |y| \le a$) without slit, or with a narrow
radial slit of width $\Delta\ll a$ extending along the $x$ axis
from $x=a_1$ to $x=a$ (Fig.~1). The current $I$ circulating
around the hole is forced to zero either ``artificially''
by putting the stream function $g(x,y)=0$ everywhere outside
the superconducting film, i.e., also in the hole and slit, or,
naturally, by cutting a slit that makes $I=0$. A uniform
magnetic field $B_a = \mu_0 H_a$ applied along $z$ is screened
inside the film (or partly screened if the 2D penetration depth
is $\Lambda = \lambda^2 /d > 0$ for thickness $d < \lambda$)
by a sheet current ${\bf J} = -{\bf\hat z} \times \nabla g =
 \nabla \times ({\bf\hat z}g) = ( \partial g / \partial y,\,
 -\partial g / \partial x )$ that circulates clockwise
near the edges and anticlockwise near the hole, see
Fig.~1 (top right). This screening current causes a
magnetic field in the hole (and in the slit) that can
be much higher than $H_a$, see Figs.~4 and 5.

  This field enhancement (or flux focussing) is plotted versus
the relative hole size $a_1/a$ in Fig.~6. For squares with
no slit our numerics yields for small holes with $a_1/a \ll 1$
and ideal screening ($\Lambda=0$) for the central field
$B(0) = \mu_0 H_z(0,0) \approx 0.40 B_a\, a/a_1$,
and for the magnetic flux in the hole
$\Phi \approx 0.80 (4a_1^2 B_a)\,a/a_1$, i.e., both values
diverge as $1/a_1$ when $a_1 \to 0$. This result is very
similar to the flux focussing in circular rings depicted
in Figs.~15 and 16 of Ref.~\onlinecite{4}.
In particular, compared to a ring with radius $a$ and
hole radius $a_1$, the limit $\Lambda = 0$ (where the flux
$\Phi$ equals the fluxoid $\Phi_{\rm f}$) is almost identical
both for $B(0)/B_a$ and for the trapped flux plotted as
the effective area $A_{\rm eff} =\Phi_{\rm f} /B_a$ referred to
the hole area. Namely, for small holes at $\Lambda=0$
one has for squares
$A_{\rm eff}/4a_1^2 \approx 0.80 a/a_1$ and for
circular rings $A_{\rm eff}/\pi a_1^2 = (8/\pi^2) a/a_1=
0.81 a/a_1$ (exact result \cite{3,4,18}).

  The corresponding results for the fluxoid $\Phi_{\rm f}$
in squares without slit are shown in Fig.~7. As expected,
$\Phi_{\rm f}$ is independent of the integration path
around the hole, which was chosen as any concentric
square of half width between $a_1$ and $a$. Again,
these $\Phi_{\rm f}$ are very similar to those for circular
rings shown in Fig.~14 of Ref.~\onlinecite{4}. For
$\Lambda=0$ (where $\Phi_{\rm f} = \Phi$) they agree closely,
as discussed above. But even for large $\Lambda \gg a$
and all hole sizes one has for rings \cite{4}
$A_{\rm eff} =2(a^2 -a_1^2)/\ln(a/a_1)$, which is also a
good approximation for the square, see the dots in Fig.~7.

  Our 2D method allows us to check this approximation
of slit-free flux focussing by considering squares or
rings with slit. The results for the square with slit
are depicted in Fig.~6 [$B(0)$ and $\Phi$ plotted
versus $1-a_1/a$] and Fig.~8 ($\Phi_{\rm f}$ for two different
integration paths). One can see that for large holes
these realistic $B(0)$, $\Phi$, and $\Phi_{\rm f}$ are similar
to those of slit-free squares. However, for small holes
$a_1/a \le 0.2$, the field enhancement is considerably
weakened by the presence of a slit, in particular for
small $\Lambda/a$. For $\Lambda=0$ the enhancement
factor does no longer diverge as $1/a_1$ but tends
to saturate (or possibly diverges very weakly,
as one over some logarithm, as is the case for finite
$\Lambda$ in the absence of a slit). This means
that the slit changes the screening currents near
small holes considerably and thus reduces the
field enhancement. As can be seen from the curves in
Fig.~6, the presence of a slit has qualitatively the
same effect as an increased value of $\Lambda$.
This finding applies even though the width $\Delta$ of
our slit was very small, $\Delta / a \approx 1/500$
and less.

  By the same token, the presence of a slit changes
the fluxoid $\Phi_{\rm f}$, Fig.~8. Moreover, the fluxoid now
depends on the integration path. Since this path has to
run inside the superconductor, one has to bridge the slit
by a narrow superconducting bridge. The current through
this bridge by definition is $I=0$ in this flux-focussing
case. It turns out that the resulting $\Phi_{\rm f}$ depends
on the position of this bridge. In Fig.~8 the two extreme
cases are shown when this bridge is chosen at $x=a_1$
(where the slit emerges from the hole) and at $x=a$ (where
the slit exits the square). For large holes these two
choices yield similar $\Phi_{\rm f}$, but for small holes the
large integration path yields a larger fluxoid than the
small path.

   One reason for this difference is that the fluxoid for
the large path includes the magnetic flux $\Phi^{\rm slit}$
inside the slit. For $\Lambda=0$ one can show that
$\Phi^{\rm slit}$ is {\em very large} and almost does
not decrease when the slit width $\Delta$ is decreased.
From the simplified model of two long parallel strips
(length $l \gg 2a$) with borders at $y=\pm a$ and
$y = \pm \Delta/2$ in perpendicular field $B_a$, one finds
for narrow slits the trapped flux \cite{7,19}
  \begin{equation}  
  \Phi^{\rm slit}  = \pi a l B_a /\ln (8a/\Delta)
  ~~~{\rm for}~~ \Delta \ll a \,.
  \end{equation}
For our squares with small hole we put the slit length
$l\approx a$ and estimate the flux in the slit as
$\Phi^{\rm slit}  = \pi a^2 B_a /\ln (8a/\Delta)$, which
depends weakly on the slit width $\Delta$. Comparing this
with the flux in small holes from above,
$\Phi^{\rm hole}  = (8/\pi) a_1 a B_a $, we get the ratio
  \begin{equation}  
  { \Phi^{\rm slit} \over \Phi^{\rm hole} } =
  {\pi^2 \over 8 \ln(8a/\Delta) }\,{a\over a_1}
  \approx 0.15\, {a\over a_1}
  \end{equation}
when $\Delta/a$ is of the order $1/500$. This means, for small
holes with relative width $a_1/a < 0.15$, the magnetic flux
even in a very narrow slit {\em exceeds} the flux in the
hole. This finding explains why for small holes our
flux-focussing results  for slit-free and slitted
disks differ considerably while they agree for large holes.

  To check this further, we computed the magnetic flux
$\Phi$ and the fluxoid $\Phi_{\rm f}$ for a square with no hole,
but with a narrow radial slit ranging from $x=0$ to $x=a$
on the $x$ axis. We bridge this slit by a narrow
superconducting bridge
centered at $x=a_2$, $0 < a_2 < a$, see Fig.~9. The contour
within which $\Phi$ and $\Phi_{\rm f}$ are calculated is a concentric
square passing through this bridge at $x=a_2$. We consider
the case of flux-focussing ($H_a>0$, $I=0$), thus the current
through the bridge is zero. In Fig.~10 the resulting $\Phi$
and $\Phi_{\rm f}$ are plotted versus $a_2/a$ for $\Lambda/a$ = 0,
0.001, 0.003, 0.01, 0.03, 0.1, 0.3, 1, and $\infty$.
Note that the scales differ by a factor of 4. As expected,
for $\Lambda=0$ (ideal screening) one has $\Phi_{\rm f} =\Phi$, and
for $\Lambda/a \gg 1$ (full penetration) $\Phi = 4a_2^2 B_a$.
Interestingly, For $\Lambda/a \ll 1$, one has approximately
$\Phi_{\rm f} \approx \Phi \approx (c a_2/a) 4a^2 B_a$, where the
constant $c\approx 0.08$ slightly depends on the slit width.
This proportionality of the flux to the slit length within the
square path, confirms that the magnetic flux in the narrow
slit is approximately proportional to its length, and that
Eq.~(31) (derived for a long double strip) is a good
approximation for our radial slit in the square.

\section{Conclusion}  

    In this paper I  presented a method that allows
to calculate the 2D distributions of the sheet current
${\bf J}(x,y)$ and magnetic field component $H_z(x,y)$
(and, of course, the full 3D magnetic field) for thin flat
superconductors of arbitrary shape. If the film thickness
is $d < \lambda$ (the London depth), our method accounts for
the 2D magnetic penetration depth $\Lambda = \lambda^2 /d$,
which may have any value, $0 \le \Lambda < \infty$. The
sheet current is expressed by the scalar potential (or stream
function) $g(x,y)$, for which we list 10 useful properties
in Sec.~II A. The statics and dynamics of superconductors in
the Meissner state, or with pinned and depinning vortices
described by a complex or nonlinear resistivity, can be
calculated. It is shown how this is generalized to multiply
connected film shapes, e.g., squares or disks with a hole
or closed slot, or with several such holes, slots, slits.
For individual 2D vortices in the film we give their mutual
interaction and self energy, which both depend on the
specimen shape. The coupling of part of the magnetic flux
of a moving vortex into a hole or slit is expressed in terms
of the stream function $g(x,y)$, which is computed and
depicted for some basic examples.

   If a slitted square or circular disk is used as a SQUID,
the applied magnetic field and applied currents induce a
signal across the weak link that bridges the slit, and the
moving vortices cause a noisy signal.\cite{1}  The SQUID
signal in principle can be calculated by our 2D method,
see Ref.~\onlinecite{7} for a detailed theory of such SQUIDs
and for the 1D problem of a long double strip that models
a linear SQUID.

  As a useful example for application of our 2D method we
consider in Sec.~III the phenomenon of flux focussing, which
occurs when the total current circulating in a square or
circular disk around a central hole is zero, $I=0$. We compare
the  ``ideal case'' when the disk has no slit ($I=0$ can then
be achieved by appropriate magnetic history) with the
realistic situation where a radial slit forces $I=0$.
We find good agreement for large holes, but for squares with
small central hole and radial slit, flux focussing is reduced
by this slit. More applications will be published.

\acknowledgments
Helpful discussions with J.\ R.\ Clem, D.~Koelle,
G.\ P.\ Mikitik, Rinke Wijngaarden, and E. Zeldov are
gratefully acknowledged.
This work was supported in part by the German Israeli
Research Grant Agreement (GIF) No G-705-50.14/01.

\appendix
\section{Numerical tricks}  

  For the computation of the stream function
$g({\bf r}) = g(x,y)$ with symmetry $g(x,-y) = g(x,y)$,
$-a \le x \le a$, $0\le y \le b$, a 2D grid of $n$ points
${\bf r}_k = (x_k, y_k)$  with weights $w_k$, $k=1 \dots n$,
is chosen that covers the basic area $2a \times b$ (upper half
of the rectangular film $2a \times 2b$). For simplicity we
chose a rectangular grid ${\bf r}_{ij} = (x_i, y_j)$ with
$i=1 \dots 2n_x$ and $j=1 \dots n_y$, e.g., the equidistant
grid $x_i =-a +(i -{1\over2})a/n_x$, $y_j = (j -{1\over2})b/n_y$
with constant weights $w = 2ab/n$, $n=2n_xn_y$.
A better choice is a nonequidistant grid that is denser near
the edges of the thin film. For example, for a rectangular
film with rectangular hole with borders at $x = \pm a_1$,
$y = \pm b_1$ and a narrow slit along $y=0$ (see Fig.~1),
a possible choice for the $y_j$ and weights $w_{yj}$
(and correspondingly for the $x_i$ and $w_{xi}$) is
$y_j =b_1 f(v_j)$, $w_{yj} =b_1 f'(v_j)/n_{y1}$,
$v_j =(j-{1\over2})/n_{y1}$ for $j=1 \dots n_{y1}$,
and
$y_j=b_1+(b-b_1) f(v_j)$, $w_{yj} =(b-b_1) f'(v_j)/(n_y-n_{y1})$,
$v_j=(j-n_{y1}-{1\over2})/(n_y-n_{y1})$ for $j=n_{y1}+1\dots n_y$,
with any function $f(v)$ defined in $0 \le v \le 1$ and having
a derivative $f'(v) = df/dv$ that at $v=0$ and $v=1$ is zero or
small. We choose, e.g.,
   \begin{eqnarray}  
    f(v)  &=& (3v^2 -2v^3) \,c +(1-c)\, v \,, \nonumber \\
    f'(v) &=& 6\,v(1-v) \,c     +1-c \,,
   \end{eqnarray}
with $0 \le c \le 1$; $c=0$ means equidistant $y_j$ with
constant weights $w_{yj}$, and $c=1$ (highest accuracy)
means that the distances
$y_{j+1} -y_j$ and weights $w_{yj}$ vanish linearly at
$y=0$, $y=b_1$, and $y=b$. One can show that
$\sum_j w_{yj} \varphi(y_j) \approx \int_0^b \varphi(y) dy$
for any sufficiently smooth function $\varphi(y)$.

   For our 2D grid ${\bf r}_{ij} = (x_i, y_j)$ the weights
are $w_{ij} = w_{xi} w_{yj}$. This 2D counting of grid points
is required for graphics and for computing derivatives,
e.g., $\partial g(x,y)/\partial x$ and
$\partial^2 g(x,y)/\partial x^2$. However, for computation of
2D integrals or for matrix operations, a 1D counting of the
grid points is required: ${\bf r}_{ij} ={\bf r}_k =(x_k,y_k)$,
$w_{ij} = w_k$, $k=i +2n_x (j-1) = 1 \dots n$, $n= 2 n_x n_y$,
and any 2D function like $g(x,y)$ is represented as a vector
$g_k = g(x_k, y_k)$. The magnetic moment of the film, Sec.~2A,
is then $m =\int g({\bf r})d^2r \approx \sum_k g_k w_k$.

   The magnetic flux through a closed rectangular path
running between the grid points (along $x$ and $y$ values
obtained from the above formulae for $x_i$, $y_i$ by using
half-integer values for $i$, $j$, e.g., $j=7/2$) is then
 $\Phi = \mu_0 \sum'_k w_k H_z(x_k,y_k)$ where the sum is
over all points $k$ inside this loop.

   Particular attention requires the computation of the
Laplacian  $\nabla^2$ acting on $g(x,y)$ [e.g., in Eq.~(13)],
and computed by multiplication by a matrix $\nabla^2_{kl}$
[e.g., in Eq.~(15)]. This operator should contain the
information that $g(x,y)=0$ outside and on the outer edges
of the rectangular film $|x| \le a$, $|y| \le b$, and that
at $y=0$ one has $\partial g(x,y) /\partial y =0$ due to the
symmetry $g(x,-y) = g(x,y)$. A 2D method that in principle
applies to any 2D grid ${\bf r}_k = (x_k, y_k)$ computes
$\nabla^2_{kl}$ as the inverse of the Green function
$G({\bf r, r}')$  satisfying
$\nabla^2 G({\bf r, r}') = \delta({\bf r - r}')$ and the
conditions that $G({\bf r, r}') = 0$ for ${\bf r}$ on the
outer edge of the rectangle $2a \times 2b$ and having even
symmetry with respect to $y$. This $G({\bf r, r}')$ may be
expressed by an infinite sum, with alternating signs, of
functions $\ln | {\bf r-r}' -{\bf R}_{mn} | /(4\pi)$, where
the ${\bf R}_{mn}$ are the vectors of a rectangular lattice
with spacings $4a$ and $2b$. The resulting matrix indeed
works, however, it is less accurate (and takes much more
computation time) than the simple 1D method of computing
 $\nabla^2 g = \partial^2 g / \partial x^2 +
 \partial^2 g / \partial y^2$ from $g(x_i, y_j)$ and the
values $g( x_{i\pm 1}, y_{j\pm 1})$ at the four
neighboring points. With our nonequidistant grid we need
for this the formula for $f''(x_i) = \partial^2 f /
\partial x^2$ at $x=x_i$. Writing
$f(x_{i-1}) = f_-$, $f(x_i) = f_0$, $f(x_{i+1}) = f_+$,
$h_1 =x_i -x_{i-1} >0$, $h_2 =x_{i+1} -x_i >0$, one has
   \begin{eqnarray}  
   f''(x_0) \approx f_- {2/h_1 \over h_1 +h_2}
  -f_0 {2 \over h_1 h_2} +f_+ {2/h_2 \over h_1 +h_2} \,.
   \end{eqnarray}
The boundary and symmetry conditions for $g(x,y)$ allow to
define the required values of $g$ on the grid lines lying
one grid spacing outside the basic area $-a \le x \le a$,
$0\le y \le b$: $x_0 = -2a +x_1$, $x_{n_x+1} = 2a -x_{n_x}$,
$y_0 =y_{-1}$, $y_{n_y +1} =2b-y_{n_y}$,
$g(x_0, y_j)=g(x_{n_x+1}, y_j) = g(x_i, y_{n_y+1}) =0$,
$g(x_i, y_0) = g(x_i, y_1)$.

\end{document}